\documentstyle[12pt]{article}

\begin{document}

\author{Minu Joy$^{\dagger }$ and V. C. Kuriakose$^{\ddagger }$ \\
Department of Physics, Cochin University of Science and Technology, \\
India-682 022}
\title{Phase transitions and bubble nucleations for a $\phi ^6$ model in curved
spacetime}
\maketitle

\begin{abstract}
Condsidering a massive self-interacting $\phi ^6$ scalar field coupled
arbitrarily to a (2+1) dimensional Bianchi type-I spacetime, we evaluate the
one-loop effective potential. It is found that $\phi ^6$ potential can be
regularized in (2+1) dimensional curved spacetime. A finite expression for
the energy-momentum tensor is obtained for this model. Evaluating the finite
temperature effective potential, the temperature dependence of phase
transitions is studied. The crucial dependence of the phase transitions on
the spacetime curvature and on the coupling to gravity are also verified. We
also discuss the nucleation of bubbles in a $\phi ^6$ model. It is found
that there exists an exact solution for the damped motion of the bubble in
the thin wall regime.

\bigskip\ $\dagger $e-mail: minujoy@cusat.ac.in

$^{\ddagger }$e-mail: vck@cusat.ac.in

\bigskip\ 

PACS number(s): 04.62.+v, 11.10.Gh, 11.10.Wx, 11.15.Ex
\end{abstract}

\baselineskip 28pt

Gravity in (2+1) dimensions [1] exhibits novel features of interest. There
are several important differences between the three and four dimensional
problems. First of all the divergence in the gravitation action induced by
scalar loops in 4 dimensions can, by power counting be proportional to 1, R,
R$^2$, R$_{\mu \nu }$ R$^{\mu \nu }$ and R$_{\alpha \beta \gamma \delta }$R$%
^{\alpha \beta \gamma \delta }($or suitable combinations of them.). In three
dimensions the situation is simplified, as the only candidates are 1 and R
[2]. Over the past two decades (2 + 1) dimensional gravity has become an
active field of research, drawing insights from general relativity. The task
of quantizing general relativity remains one of the outstanding problems of
theoretical physics. General relativity is a geometric theory of spacetime,
and quantising gravity means quantising spacetime itself. Ordinary quantum
field theory is local, but the fundamental physical observables of quantum
gravity are necessarily nonlocal. Ordinary quantum field theory takes
causality as a fundamental postulate, but in quantum gravity the spacetime
geometry and thus the light cones and the causal structure, are themselves
subject to quantum fluctuations. Again, perturbative quantum field theory
depends on the existence of a smooth, approximately flat background, but
there is no reason to believe that the short-distance limit of quantum
gravity even resembles a smooth manifold. Faced with these problems, it is
natural to look for simpler models that share the important conceptual
features of general relativity while avoiding some of the conceptual
difficulties. General relativity in (2 + 1) dimensions is one such model. As
a generally covariant theory of spacetime geometry, (2 + 1) dimensional
gravity has the same conceptual foundations as relativistic (3 + 1)
dimensional general relativity, and many of the fundamental issues of
quantum gravity carry over to the lower dimensional setting. At the same
time, the (2 + 1) dimensional model is vastly simpler, and one can actually
write down candidates for a quantum gravity [1].

Another important feature of the conformally invariant scalar theory in
three dimensions is that its $\phi ^6$ coupling can in principle, induce a
divergence in the four-point Green's functions neccessitating a $\phi ^4\ $%
coupling, which is not conformally invariant. There is no analogue of this
possibility in four dimensions, the renormalisable $\phi ^4$ coupling is
conformally invariant and cannot generate any divergence that corresponds to
a nonconformally invariant coupling [3].

Quantum fields [4] have profound influence on the dynamical behaviour of the
early universe [5-9]. In the present work we discuss the finite temperature
phase transitions in a (2+1) dimensional curved spacetime for $\phi ^6$
model. In Ref. [9] we have obtained a finite expression for the one-loop
effective potential [10] using the $\phi ^6$ model in a (3+1) dimensional
Bianchi type-I spacetime.

Consider a massive{\sl \ } self interacting{\sl \ } scalar field{\sl \ $\phi 
$ }coupled{\sl \ }arbitrarily to the gravitational back ground and described
by the Lagrangian density $\pounds $, 
\begin{equation}
\label{1}\pounds =\sqrt{-g}\left\{ \frac 12[g^{\mu \nu }\partial _\mu \phi
\partial _\nu \phi -\xi R\phi ^2]-\frac 12\lambda ^2\phi ^2(\phi
^2-m/\lambda )^2\right\} 
\end{equation}
{\sl \ } The equation of motion associated with the Lagrangian(1) is,\ 
\begin{equation}
\label{2}g^{\mu \nu }\nabla _\mu \nabla _\nu \phi +(m^2+\xi R)\phi -4\kappa
\phi ^3+3\lambda ^2\phi ^5=0 
\end{equation}
{\sl \ }in which we put $m$$\lambda =\kappa .$ We can write $\phi =\phi
_c+\phi _q$ {\sl \ }where $\phi _c$\ is the classical field and $\phi _q$\
is\ a quantum field with vanishing vacuum expectation value, \\$<\phi _q>$ $%
=0.\ $ The field equation for the classical field $\phi _c$ is, 
\begin{equation}
\label{3}
\begin{array}{c}
g^{\mu \nu }\nabla _\mu \nabla _\nu \phi _c+[(m_r^2+\delta m^2)+(\xi
_r+\delta \xi )R]\phi _c-4(\kappa _r+\delta \kappa )\phi _c^3-12(\kappa
_r+\delta \kappa )\phi _c<\phi _q^2> \\  
\\ 
+\ 3(\lambda _r^2+\delta \lambda ^2)\phi _c^5+30(\lambda _r^2+\delta \lambda
^2)\phi _c^3<\phi _q^2>+15(\lambda _r^2+\delta \lambda ^2)\phi _c<\phi
_q^4>\ =0 
\end{array}
\end{equation}
where the bare parameters m,$\xi ,\kappa $ and $\lambda $ are replaced by
the renormalised terms [9]. To the one loop quantum effect, the field
equation for the quantum field $\phi _q$ is, 
\begin{equation}
\label{4}g^{\mu \nu }\nabla _\mu \nabla _\nu \phi _q+(m_r^2+\xi R)\phi
_q-12\kappa _r\phi _c^2\phi _q+15\lambda _r^2\phi _c^4\phi _q=0 
\end{equation}
The effective potential V$_{eff}\ $is given by, 
\begin{equation}
\label{5}
\begin{array}{l}
V_{eff}=\frac 12[(m_r^2+\delta m^2)+(\xi _r+\delta \xi )R][\phi _c^2+<\phi
_q^2>]-(\kappa _r+\delta \kappa )\phi _c^4 \\  
\\ 
\hspace{1.2cm}-6(\kappa _r+\delta \kappa )\phi _c^2<\phi _q^2>-(\kappa
_r+\delta \kappa )<\phi _q^4>+\frac 12(\lambda _r^2+\delta \lambda ^2)\phi
_c^6 \\  \\ 
\hspace{1.2cm}+\frac{15}2(\lambda _r^2+\delta \lambda ^2)\phi _c^4<\phi
_q^2>+\frac{15}2(\lambda _r^2+\delta \lambda ^2)\phi _c^2<\phi _q^4>+\frac
12(\lambda _r^2+\delta \lambda ^2)<\phi _q^6> 
\end{array}
\end{equation}
To make V$_{eff}$ finite, the following renormalisation conditions are used, 
\begin{equation}
\label{6}
\begin{array}{c}
m_r^2=\left( 
\frac{\partial ^2V_{eff}}{\partial \phi _c^2}\right) _{\phi _c=R=0}, 
\hspace{1.1cm}\xi _r=\left( \frac{\partial ^3V_{eff}}{\partial R\partial
\phi _c^2}\right) _{\phi _c=R=0}, \\  \\ 
\kappa _r=\left( \frac{\partial ^4V_{eff}}{\partial \phi _c^4}\right) _{\phi
_c{\ }=R=0},\hspace{1.1cm}{\sl \lambda _r^2=\left( \frac{\partial ^6V_{eff}}{%
\partial \phi _c^6}\right) }_{\phi _c=R=0} 
\end{array}
\end{equation}
\ To evaluate $<\phi _q^2>,$ $<\phi _q^4>,$ and $<\phi _q^6>$ we adopt the
canonical quantisation relations: 
\begin{equation}
\label{7}[\phi _q(t,x),\phi _q(t,y)]=[\pi _q(t,x),\pi _q(t,y)]=0;%
\hspace{1.3cm}[\phi _q(t,x),\pi _q(t,y)]=i\delta ^3(x-y) 
\end{equation}
where the conjugate momentum $\pi _q$ is defined by $\pi _q=\frac{\partial
\pounds }{\partial (\partial _i\phi )}$ .

We consider a (2+1) dimensional Bianchi type-I spacetime with small
anisotropy which has the line element 
\begin{equation}
\label{8}ds^2=C(\eta )d\eta ^2-a_1^2(\eta )dx^2-a_2^2(\eta )dy^2,%
\hspace{1.cm}C=a_1a_2 
\end{equation}
In this model the mode function of the quantum field $\phi _q$ can be
written in the separated form as u$_k=C^{-1/4}(2\pi )^{-1}\exp (i\kappa
.x)\chi _k(\eta ).$ The wave equation Eq. (4) will then lead to 
\begin{equation}
\label{9}\stackrel{\cdot \cdot }{\chi }+\left\{ C\left[ m_r^2+(\xi _r-\frac
18)R-12\kappa _r\phi _c^2+15\lambda _r^2\phi _c^4+\sum\limits_i\frac{k_i^2}{%
a_i^2}\right] +Q\right\} \chi _k=0 
\end{equation}
where the spacetime curvature function R and the anisotropic function Q are 
\begin{equation}
\label{10}R=8C^{-1}(\stackrel{\bullet }{H}+H^2+Q),\hspace{.6cm}%
H=\sum\limits_ih_i,\hspace{.6cm}h_i=\frac{\stackrel{\cdot }{a_i}}{a_i},%
\hspace{.6cm}Q=\frac 1{64}\sum\limits_{i<j}(h_i-h_j)^2 
\end{equation}
When the metric is slowly varying Eq. (9) possesses WKB approximation
solution: 
\begin{equation}
\label{11}\chi _k=(2W_k)^{-\frac 12}\exp (-i\int d\eta W_k) 
\end{equation}
where,$\hspace{1.5cm}W_k=\left\{ C\left[ m_r^2+(\xi _r-\frac 18)R-12\kappa
_r\phi _c^2+15\lambda _r^2\phi _c^4+\sum\limits_i\frac{k_i^2}{a_i^2}\right]
+Q\right\} ^{\frac 12}$

Using the above solution we get: 
\begin{equation}
\label{12}
\begin{array}{l}
<\phi _q^2>=\frac 1{8\pi ^2C(\eta )}\int d^2k\left[ m_r^2+(\xi _r-\frac
18)R-12\kappa _r\phi _c^2+15\lambda _r^2\phi _c^4+\sum\limits_i 
\frac{k_i^2}{a_i^2}+\frac QC\right] ^{-1/2} \\  \\ 
\hspace{1.3cm}=\frac 1{16\pi }\left[ \Lambda +\frac{(m_r^2+(\xi _r-\frac
18)R-12\kappa _r\phi _c^2+15\lambda _r^2\phi _c^4+\frac QC)}{2\Lambda }%
-(m_r^2+(\xi _r-\frac 18)R-12\kappa _r\phi _c^2+15\lambda _r^2\phi
_c^4+\frac QC)^{1/2}\right] \\  
\end{array}
\end{equation}
and similarly, 
\begin{equation}
\label{13}
\begin{array}{l}
<\phi _q^4>=\frac 1{128\pi ^3C}\log \left[ 1+ 
\frac{\Lambda ^2}{(m_r^2+(\xi _r-\frac 18)R-12\kappa _r\phi _c^2+15\lambda
_r^2\phi _c^4+\frac QC)}\right] \\  
\end{array}
\end{equation}
where we have introduced a momentum cut-off $\Lambda $ to regularise the
k-integration. From the renormalisation conditions given by Eq. (6) the
renormalisation counter terms are evaluated and substituting these terms we
find $\frac{\partial V_{eff}}{\partial \phi _c}$ obtained from Eq. (5) as, 
\begin{equation}
\label{14}
\begin{array}{l}
\frac{\partial V_{eff}}{\partial \phi _c}=(m_r^2+\xi _rR)\phi _c-\left[ 
\frac{3\lambda _r^2[(m_r^2+\frac QC)-A]}{4B}\right] \phi \\  \\  
\\ 
\hspace{1.5cm}-\left[ \frac{D(m_r^2+\frac QC)\log [\frac{(m_r^2+\frac QC)}A]%
}{2\pi CEB}\right] \phi _c+\left[ \frac{-3\lambda _r^2}{8B}+\frac D{2\pi
CEB}\right] (\xi _r-\frac 18)R\phi _c \\  \\ 
\hspace{1.5cm}-\frac{4(m_r^2+\frac QC)^{1/2}}B\left[ \pi \lambda _r^2+\frac{%
2D(m_r^2+\frac QC)^{1/2}A^{1/2}}E\right] \phi _c^3+\frac{64D(m_r^2+\frac QC)%
}{5EB}\phi _c^5 \\  
\end{array}
\end{equation}
where, 
\begin{equation}
\label{15}
\begin{array}{l}
A=(m_r^2+(\xi _r-\frac 18)R-12\kappa _r\phi _c^2+15\lambda _r^2\phi
_c^4+\frac QC), 
\hspace{.6cm}B=\left[ \frac{-1350\lambda _r^2}4+\frac{3420\kappa _r^2}{%
2(m_r^2+\frac QC)}\right] , \\  \\ 
D=\left[ \lambda _r^2-\frac{54\kappa _r\lambda _r^2}{4\pi (m_r^2+\frac
QC)^{1/2}}+\frac{135\kappa _r^2}{2\pi (m_r^2+\frac QC)^{3/2}}\right] and%
\hspace{.4cm}E=\left[ -(m_r^2+\frac QC)^{1/2}+\frac{3\kappa _r}{4\pi
(m_r^2+\frac QC)}\right] 
\end{array}
\end{equation}

Thus it is clear that we can obtain a finite expression for the one loop
effective potential for the $\phi ^6$ model in (2+1) dimensional Bianchi
Type I spacetime. In a previous work [9] we have shown that $\phi ^6$
potential can be regularised in (3+1) dimensional curved spacetime. In the
present work using the momentum cutoff technique we obtain a divergenceless
expression for the $\phi ^6$ potential in a (2+1) dimensional Bianchi type-I
background spacetime.

While constructing a theory of the interaction between quantised matter
fields and a classical gravitational field one has to identify the
energy-momentum tensor of the quantised fields which acts as the source of
the gravitational field. The energy-momentum tensor [4] for the present $%
\phi ^6$ field is 
\begin{equation}
\label{16}
\begin{array}{l}
T_{\mu \nu }=(1-2\xi )\partial _\mu \phi \partial _\nu \phi +(2\xi -\frac
12)g_{_{\mu \nu }}\partial _\alpha \phi \partial ^\alpha \phi -2\xi \phi
\bigtriangledown _\mu \bigtriangledown _\nu \phi +2\xi g_{_{\mu \nu }}\phi
\Box \phi -\xi G_{\mu \nu }\phi ^2 \\  
\\ 
\hspace{1.5cm}+(\frac{m^2}2)g_{_{\mu \nu }}\phi ^2-2\kappa g_{_{\mu \nu
}}\phi ^4+\frac 32\lambda ^2g_{_{\mu \nu }}\phi ^6 
\end{array}
\end{equation}
where $G_{\mu \nu }$ is the Einstein tensor. The expection value of the
energy-momentum tensor can be broken into classical and quantum parts. The
energy-momentum tensor for the classical part is obtained by substituting $%
\phi _c$ for $\phi $ in Eq. (16). The $\eta \eta $ component of the
classical renormalized energy-momentum tensor is given by, 
\begin{equation}
\label{17}\langle T_\eta ^\eta \rangle ^C=\frac 1{2C}\phi _c^{^{\prime
}}\phi _c^{^{\prime }}+2\xi \frac{C^{^{\prime }}}{C^2}\phi _c\phi
_c^{^{\prime }}+\frac{3\xi }C(\frac{C^{^{\prime }}}C+k)\phi _c^2+\frac{m^2}
2\phi _c^2-2\kappa \phi _c^4+\frac 32\lambda ^2\phi _c^4 
\end{equation}
where k=+1,0,-1 corresponds to the case of positive, zero or negative
spatial curvature respectively. The quantum part of is $\langle T_{_{\mu \nu
}}\rangle $ is 
\begin{equation}
\label{18}
\begin{array}{l}
\langle T_{_{\mu \nu }}\rangle ^Q=(1-2\xi )\langle \partial _\mu \phi
_q\partial _\nu \phi _q\rangle +(2\xi -\frac 12)g_{_{\mu \nu }}\langle
\partial _\alpha \phi _q\partial ^\alpha \phi _q\rangle -2\xi \langle \phi
_q\bigtriangledown _\mu \bigtriangledown _\nu \phi _q\rangle \\  
\\ 
\hspace{1.6cm}+2\xi g_{_{\mu \nu }}\langle \phi _q\Box \phi _q\rangle -\xi
G_{\mu \nu }\langle \phi _q^2\rangle +(\frac{m^2}2)g_{_{\mu \nu }}\langle
\phi _q^2\rangle -12\kappa g_{_{\mu \nu }}\phi _c^2\langle \phi _q^2\rangle
\\  \\ 
\hspace{1.6cm}+\frac{45}2\lambda ^2g_{_{\mu \nu }}\phi _c^4\langle \phi
_q^2\rangle +\frac{45}2\lambda ^2g_{_{\mu \nu }}\phi _c^2\langle \phi
_q^4\rangle 
\end{array}
\end{equation}
To obtain the physically finite energy-momentum tensor of the system we can
regularize the theory [11]. The finite expression for the expectation value
of the quantum energy-momentum tensor is obtained as,

\begin{equation}
\label{19}
\begin{array}{l}
\langle T_\eta ^\eta \rangle ^Q= 
\frac{C^{\prime 2}[3A-\frac QC]}{768\pi C^4A^{3/2}}-\frac 1{48\pi }A^{3/2}- 
\frac{C^{\prime 2}A^{1/2}}{256\pi C^3}-\frac{C^{\prime 2}(A+\frac QC)}{
128\pi C^3A^{1/2}} \\  \\ 
\hspace{1.6cm}+\frac{C^{\prime 2}}{16\pi C^3}\left\{ \xi _r-\frac{3(\xi
_r-\frac 18)\lambda _r^2}{8B}+\frac{D(\xi _r-\frac 18)}{2CBE}\right\} \frac{
(A+\frac QC)}{A^{1/2}} \\  \\ 
\hspace{1.6cm}+\frac{C^{^{\prime }}}{16\pi C^2}[\frac{C^{\prime }}C-3 
]\left\{ \xi _r-\frac{3(\xi _r-\frac 18)\lambda _r^2}{8B}+\frac{D(\xi
_r-\frac 18)}{2CBE}\right\} A^{1/2} \\  \\ 
\hspace{1.6cm}+\frac{3(m_r^2+\frac QC)}{4B}\left\{ \lambda _r^2-\frac{D\log
(m_r^2+\frac QC)}{\pi CE}\right\} -\frac{6D(m_r^2+\frac QC)^{3/2}}{BE} 
\end{array}
\end{equation}
where A, B and E are defined by Eq. (15). It is clear that the
energy-momentum tensor depends on the anisotropy of the spacetime.

To evaluate the finite temperature effective potential, the vaccum
expectation value is replaced by the thermal average $<\phi >_T=\sigma _T$
taken with respect to a Gibbs ensemble [12]. Considering the same Lagrangian
density as above, the zero loop effective potential is temperature
independent, 
\begin{equation}
\label{20}V_0(\sigma )=\frac 12\xi R\sigma ^2+\frac 12\lambda ^2\sigma
^2(\sigma ^2-m/\lambda )^2 
\end{equation}
The one loop approximation to finite temperature effective potential
[10,12-14] is given by,

\begin{equation}
\label{21}
\begin{array}{l}
V_1^\beta (\sigma )=\frac 1{2\beta }\sum_n\int 
\frac{d^2k}{(2\pi )^2}\ln (k^2-M^2) \\  \\ 
\hspace{1.92cm}=\frac 1{2\beta }\sum_n\int \frac{d^2k}{(2\pi )^2}\ln (\frac{
-4\pi ^2n^2}{\beta ^2}-E_M^2) 
\end{array}
\end{equation}

\begin{equation}
\label{22}where,E_M^2=k^2+M^2,\hspace{1.5cm}M^2=m^2+\xi R-12\lambda m\sigma
^2+15\lambda ^2\sigma ^4 
\end{equation}
In the high temperature limit [13] we find that 
\begin{equation}
\label{23}V_1^\beta (\sigma )=\frac 1{4\pi \beta ^3}\xi (z)-\frac{M^2}{8\pi
\beta }\ln M 
\end{equation}
where $\xi (z)$ is the Riemannian Zeta function. In the (2+1) dimensional
case also, the symmetry breaking present in this $\phi ^6$ model can be
removed if the temperature is raised above a certain value called the
critical temperature. The order parameter of the theory is temperature
dependent. The temperature dependence of finite temperature effective
potential leads to phase transitions in the early universe.

On shifting the field from $\phi $ to $\phi +${\sl $\sigma $ }in the Eq. (2) 
{\sl \ }and taking the Gibbs average of the corresponding equation we get: 
\begin{equation}
\label{24}
\begin{array}{l}
\Box \sigma _T+(m^2+\xi R)\sigma _T-4\kappa \sigma _T{}^3-12\kappa \sigma
_T<\phi ^2>-12\kappa \sigma _T^2<\phi >+15\lambda ^2\sigma _T<\phi ^4> \\  
\\ 
\hspace{1cm}+30\lambda ^2\sigma _T^2<\phi ^3>+30\lambda ^2\sigma _T^3<\phi
^2>+15\lambda ^2\sigma _T^4<\phi >+3\lambda ^2\sigma _T^5=0 
\end{array}
\end{equation}
Using the standard finite temperature Green's- function methods we can find
that in the high temperature limit, $<\phi ^2>=\frac{-m}{4\pi T}$ and from
similar calculations, $<\phi ^4>=\frac T{4\pi m},\hspace{.4cm}<\phi ^3>=0$
and $<\phi >=0.$ Thus Eq. (24) becomes 
\begin{equation}
\label{25}\Box \sigma _T+(m^2+\xi R)\sigma _T-4\kappa \sigma _T^3+\frac{
3\kappa \sigma _Tm}{\pi T}+\frac{15\lambda ^2\sigma _TT}{4\pi m}-\frac{
15\lambda ^2\sigma _T^3m}{2\pi T}+3\lambda ^2\sigma _T^5=0 
\end{equation}
Assuming that $\sigma _T$ is a constant we obtain, 
\begin{equation}
\label{26}\sigma _T\left[ (m^2+\xi R)-4\kappa \sigma _T^2+\frac{3\kappa m}{
\pi T}+\frac{15\lambda ^2T}{4\pi m}-\frac{15\lambda ^2\sigma _T^2m}{2\pi T}
+3\lambda ^2\sigma _T^4\right] =0 
\end{equation}
This equation has degenerate solutions: 
\begin{equation}
\label{27}\sigma _T=0, and\hspace{.5cm}\sigma _T^2=\frac{\left\{ (4\kappa +%
\frac{15\lambda ^2m}{2\pi T})\pm \left( 4\lambda ^2m^2-12\lambda ^2\xi R+%
\frac{24\lambda ^3m^2}{\pi T}+\frac{225\lambda ^4m^2}{4\pi ^2T^2}- \frac{%
45\lambda ^4T}{\pi m}\right) ^{\frac 12}\right\} }{6\lambda ^2} 
\end{equation}
Each solution of these equations defines a possible phase of the field
system with its characteristic excitations. On heating the field system from
absolute zero, the two branches of $\sigma _T^2$ given by the above equation
coincide at a temperature for which 
\begin{equation}
\label{28}\left( 4\lambda ^2m^2-12\lambda ^2\xi R+\frac{24\lambda ^3m^2}{\pi
T}+\frac{225\lambda ^4m^2}{4\pi ^2T^2}-\frac{45\lambda ^4T}{\pi m}\right)
^{\frac 12}=0 
\end{equation}
yielding a common value of $\sigma _T.$ The existence of the separate
branches of $\sigma _T^2$ implies that the phase transition is of first
order [12,15]. From Eq. (27) it is clear that the order parameter does not
vanish even for very high values of the temperature. Fig. 1 gives the
variation of the two branches of $\sigma _T^2$ with respect to temperature.
It is found that the two branches coincides at a particular value of T given
by Eq. (28). From the graphs it is clear that there is a discontinuity for
the variation of the order parameter with temperature, indicating a first
order phase transition. For a second order phase transition the order
parameter will be a continuous function of T and it decreases smoothly with
increasing temperature and vanishes at T$_c$ [12].

The characteristic of a first order phase transition is the existence of a
barrier between the symmetric and the broken phase [16]. The temperature
dependence of V$_{eff}$ for a first order phase transition obtained using
the present $\phi ^6$ model in the (2+1) dimensional background spacetime is
shown in Fig. 2. It is found that for T$\gg $T$_c,$ the effective potential
attains a minimum at $\sigma =0,$ which corresponds to the completely
symmetric case. When the temperature decreases, a global minimum appears at $%
\sigma =0$ and two local minima at $\sigma \neq 0,$ which shows the
existence of a barrier between the global and local minima (Fig. 2a)$.$ At
T=T$_c$, all the minima are degenerate, which implies that the symmetry is
broken (Fig. 2b). For T<T$_c$ the minima at $\sigma \neq 0$ become global
minima (Fig. 2c). If for T$\leq T_c$ the extremum at $\sigma =0 $ remains a
local minimum, there must be a barrier between the minimum at $\sigma =0$
and at $\sigma \neq 0.$ Therefore the change in $\sigma $ in going from one
phase to the other must be discontinuous, indicating a first order phase
transition [12,16].

Fig. 3 clearly shows the crucial role of scalar curvature R in determining
the fate of symmetry and the phase transions for the present model. From the
figure it is clear that the first order phase transition takes place as R
changes. It is found that for $R=0$ or $\xi =0$ the system remains in the
symmetry broken state for all values of T$\leq T_c$. As the temperature is
increased above T$_c,$ the symmetry is restored depending on the values of $%
R $ and $\xi $ also$.$ It is also found that symmetry can be restored either
by increasing the value of $R$ or by increasing the value of $\xi $ keeping
the temperature constant, even below the critical temperature. It is clear
from Fig. 4 that there is a barrier between the symmetric and broken phases.
Thus phase transition, induced by the coupling constant $\xi $ is also of
first order. This shows that the scalar-gravitational coupling and the
scalar curvature do play a crucial role in determining the nature of phase
transitions took place in the early universe.

A first order phase transition proceeds by nucleation of bubbles of broken
phase in the background of unbroken phase [12]. Bubble nucleation can have
various consequences in the early universe. The bubbles expand and
eventually collide, while new bubbles are continuously formed, until the
phase transition is completed.

While discussing the bubble collisions [17-18], one has to consider the
interaction between the bubble field and the surrounding plasma. Considering
the form of Lagrangian as in Eq. (1) for a massive{\sl \ } self interacting%
{\sl \ }complex field $\Phi ${\sl \ }coupled{\sl \ }arbitrarily to the
gravitational back ground, the from of potential is 
\begin{equation}
\label{29}V(\phi )=\frac 12\xi R\left| \Phi \right| ^2+\frac 12\lambda
^2\left| \Phi ^2\right| (\left| \Phi \right| ^2-m/\lambda )^2 
\end{equation}
with a minimum at $\left| \Phi \right| =0$ and a set of minima at $\left|
\Phi \right| =\left[ \frac{m\pm \sqrt{-\xi R}}\lambda \right] ^{1/2},$
connected by U(1) transformation and towards which the false vacuum will
decay via bubble nucleation. The equation of motion for this system is 
\begin{equation}
\label{30}\partial _\mu \partial ^\mu \Phi =-\frac{\partial V}{\partial \Phi 
} 
\end{equation}
Let us consider the minimally coupled case $\xi =0$. For the thin wall
regime [12] the approximate solution of Eq. (30) is obtained as 
\begin{equation}
\label{31}\left| \Phi \right| =\left\{ \frac m{2\lambda }\left[ \tanh \left(
m(\chi -R_0)+1\right) \right] \right\} ^{1/2} 
\end{equation}
where $R_0$ is the bubble radius at nucleation time and $\chi ^2=\mid \vec
x\mid ^2-t^2.$ Cosidering the damping effect of the surrounding plasma on
the motion of the walls we insert a frictional term in the equation of
motion , 
\begin{equation}
\label{32}\partial _\mu \partial ^\mu \Phi +\gamma \left| \dot \Phi \right|
e^{i\theta }=-\frac{\partial V}{\partial \Phi } 
\end{equation}
where $\left| \dot \Phi \right| =\frac{\partial \left| \Phi \right| }{%
\partial t},$ $\theta $ is the phase of the field and $\gamma $ stands for
the friction coefficient.

To find the solution of Eq. (32) in the thin wall limit, first we suppose
that solution for which the wall has the form of a travelling wave do exist.
Writing $\Phi $ in polar form $\Phi =\rho e^{i\theta }$ , we can rewrite Eq.
(32) and for a single bubble configuration we take the phase of the bubble $%
\theta $ to be constant. Then the equation for the modulus of the field is 
\begin{equation}
\label{33}\partial _\mu \partial ^\mu \rho +\gamma \dot \rho =-\frac{%
\partial V(\rho )}{\partial \rho } 
\end{equation}
Following Ferrera and Melfo [18] we get 
\begin{equation}
\label{34}(1-v_{ter}^2)\frac{\partial ^2\rho }{\partial r^2}=\frac{\partial
V(\rho )}{\partial \rho } 
\end{equation}
where r is the radial coordinate and $v_{ter}$ is the terminal velocity of
the bubble walls. For the present $\phi ^6$ potential the solution for Eq.
(34) is obtained as 
\begin{equation}
\label{35}\rho =\left\{ \frac m{2\lambda }\left[ \tanh \left( \frac{%
m(r-v_{ter}t-R_0)}{\sqrt{1-v_{ter}^2}}\right) +1\right] \right\} ^{1/2} 
\end{equation}
which is simply a Lorentz-contracted moving domain wall. Thus it is clear
that there exists an exact solution for the damped motion of the bubble in
the thin wall regime.

Whether or not the Universe recovers from a first order phase transition and
any relics are left behind depends upon the kinematics of bubble nucleation
and on the process of eventual transition to the new phase.

\begin{center}
{\bf ACKNOWLEDGEMENTS}
\end{center}

One of us (MJ) would like to thank UGC, N. Delhi for financial support. VCK\
acknowledges Associateship of IUCAA, Pune.

\begin{enumerate}
\item  Fig. 1 : Variation of the two branches of $\sigma _T^2$ with respect
to temperature. The two curves coincides after the temperature which
satisfies Eq. (48), where $m=0.9371$, $\lambda =0.008,$ $R=0.9$ and $\xi
=0.2 $ in the figure.

\item  Fig. 2a : The behaviour of finite temperature effective potential as
a function of $\sigma $ for fixed $m=0.9371$, $\lambda =0.008,$ $R=1.93,\xi
=0.098$ and $T=16.5$ such that ${\bf T>T}_c{\bf .}$

\item  Fig. 2b : The behaviour of finite temperature effective potential as
a function of $\sigma $ for fixed $m=0.9371$, $\lambda =0.008,$ $R=0.42,\xi
=0.02$ and $T=9$ such that ${\bf T=T}_c{\bf .}$

\item  Fig. 2c : The behaviour of finite temperature effective potential as
a function of $\sigma $ for fixed $m=0.9371$, $\lambda =0.008,$ $R=0.35,\xi
=-0.3$ and $T=5$ such that ${\bf T<T}_c{\bf .}$

\item  Fig. 3 : The behaviour of finite temperature effective potential as a
function of $\sigma $ for fixed $m=0.9371$, $\lambda =0.009,$ $\xi =0.1$ and
T$=4.$ Starting from top the curves corresponds to the following values of
the curvature : R=20,3,0.99,0.02,-0.72.

\item  Fig. 4 : The behaviour of finite temperature effective potential as a
function of $\sigma $ for fixed $m=0.9371$, $\lambda =0.009,$ $R=0.2$ and $%
T=5.$ Starting from top the curves corresponds to the following values of
the curvature : $\xi $ = 9,2,0.85,0.025,-0.35,-0.8.
\end{enumerate}


\begin{thebibliography}{99}
\bibitem{1}  Steven Carlip, Quantum Gravity in 2+1 Dimensions, Cambridge
University Press, Cambridge, England, 1998.

\bibitem{2}  L. S. Brown, J. C. Collins, Ann. Phys. NY.{\bf \ }130 (1980)
215.

\bibitem{3}  D. G. C. McKeon, G. Tsoupros, Class. Quantum Grav. 11 (1994) 73.

\bibitem{4}  N. D. Birrel, P. C. W. Davies,\ Quantum Fields in Curved Space,
Cambridge University Press, Cambridge, England, 1982.

\bibitem{5}  H. Ford, D. J. Toms, Phys. Rev. D 25 (1982){\bf \ }1510. {\bf \ 
}

\bibitem{6}  T. Inagaki, T. Muta, S. D. Odintsov, Progr. Theor. Phys. Suppl.
127{\bf \ }(1997){\bf \ }93.{\bf \ }

\bibitem{7}  W. H. Huang, Class. Quantum Grav. 10{\bf \ }(1993){\bf \ }2021.

\bibitem{8}  D. J. O'Connor, B. L. Hu, T. C. Shen, Phys. Letts. 130B{\bf \ }%
(1983){\bf \ }31.

\bibitem{9}  M. Joy, V. C. Kuriakose, Phy. Rev. D 62 (2000) 104017.

\bibitem{10}  I. L. Buchbinder, S. D. Odinstov, I. L. Shapiro, Effective
Action in Quantum Gravity,{\it \ }IOP Publishing {\it \ }Ltd, 1992.

\bibitem{11}  C. M. Paris, P. R. Anderson, S. D. Ramsey, Phys. Rev. D 61
(2000) 127501.

\bibitem{12}  A. D. Linde, Particle Physics and Inflationary Cosmology,{\it %
\ }Harwood Academic Publishers GmbH, Switzerland, 1990.

\bibitem{13}  L. Dolan, R. Jackiw, Phys. Rev. D 9 (1974) 2904.

\bibitem{14}  K. Babu Joseph, V. C. Kuriakose, J. Phys. A: Math. Gen. 15
(1982){\bf \ }2231.

\bibitem{15}  Jean-Claude Toledano, Pierre Toledano, The Landau Theory of
Phase Transitions, World Scientific, 1987, Chap. 4.

\bibitem{16}  E. W. Kolb, M. S. Turner, The Early Universe,{\it \ }%
Addison-Wesley Publishing Company, 1990.

\bibitem{17}  M. S. Turner, J. Weinberg, L. M. Widrow, Phys. Rev. D 46
(1992) 2384.

\bibitem{18}  A. Ferrera, A. Melfo, Phys. Rev. D 53 (1996) 6852.

\newpage\ {\bf Figure Captions}
\end{thebibliography}
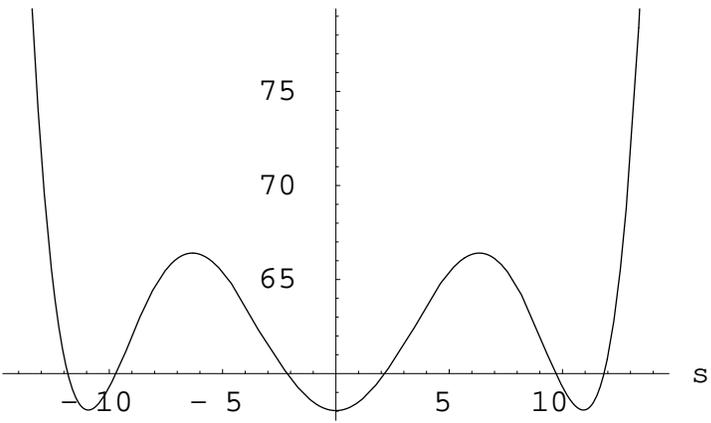
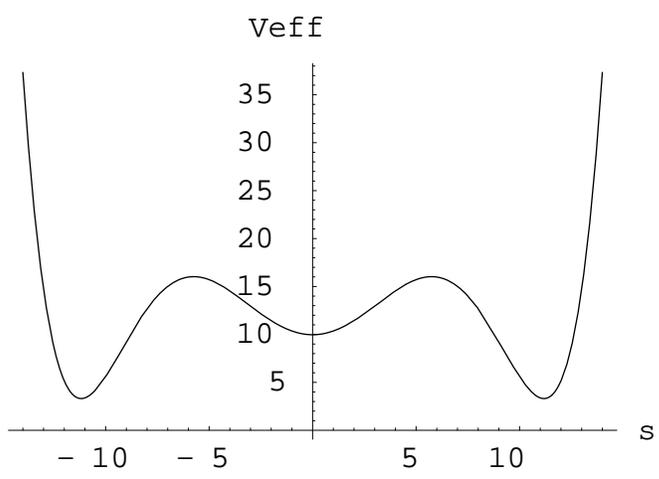
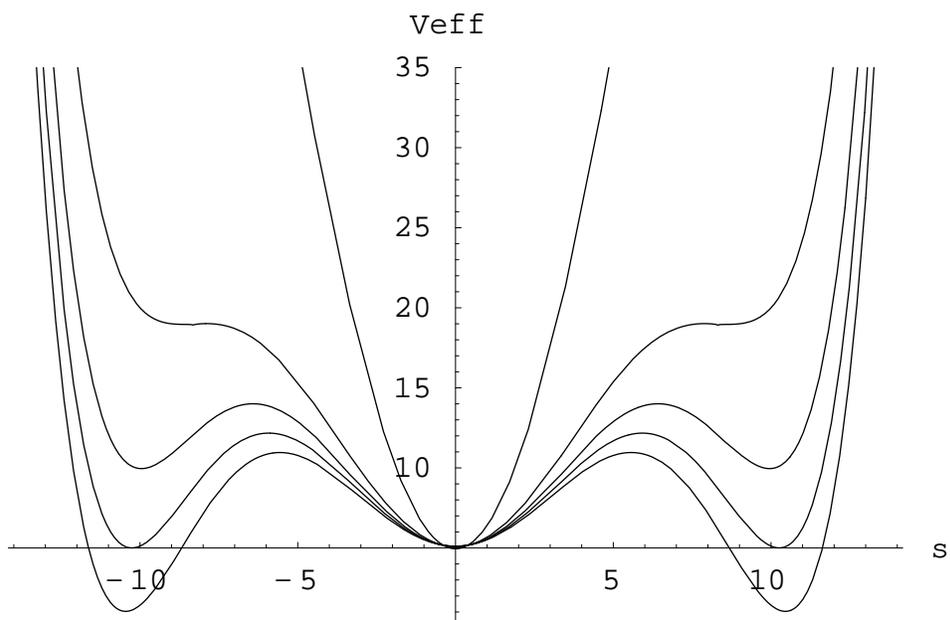
\end{document}